\documentclass[journal]{IEEEtran}
\usepackage{cite}

\ifCLASSINFOpdf
   
   \usepackage[pdftex]{graphicx}
   \usepackage{multirow}
   \usepackage{tabularew}
   \usepackage[flushleft]{threeparttable}
   
   \graphicspath{{./pdf/}{./jpeg/}{./eps/}}
  \DeclareGraphicsExtensions{ ,.jpeg,.png}
\else
   \usepackage[dvips]{graphicx}
   \graphicspath{{./eps/}}
   \DeclareGraphicsExtensions{.eps}
\fi
\usepackage{amsmath}
\usepackage{array}

\ifCLASSOPTIONcompsoc
 \usepackage[caption=false,font=normalsize,labelfont=sf,textfont=sf]{subfig}
\else
  \usepackage[caption=false,font=footnotesize]{subfig}
\fi

\usepackage{stfloats}
\usepackage{url}
  \usepackage{color}

\usepackage{url}

\begin{document}
\title{High Speed and Low Power Sensing Schemes for STT-MRAM with IPMTJs}
\author{Mesut~Atasoyu,
        Mustafa~Altun,
        Serdar~Ozoguz 
\thanks{M. Atasoyu, M. Altun and S. Ozoguz are with Department of Electronics and Communication Engineering Istanbul Technical University, Istanbul,34469 TR (e-mail: matasoyu, altunmus, ozoguz@itu.edu.tr).}}
\maketitle
\begin{abstract}
STT-MRAM with interfacial-anisotropy-type perpendicular MTJ (IPMTJ) is a powerful candidate for the low switching energy design of STT-MRAM. In the literature, the reading operation of STT-MRAM structured with IPMTJs have been not studied until this time, in our knowledge. We investigated the reading operation of STT-MRAM structured with IPMTJs. To enumerate the read operations of the NVSenseAmp have successfully been performed a 2.5X reduction in average low power and a 13X increase in average high speed compared with the previous works.
\end{abstract}
\IEEEpeerreviewmaketitle
\section{INTRODUCTION}
Recent developments in scalability of the spin-transfer torque magnetic RAM (STT-MRAM) have led to a renewed interest in a promising nonvolatile memory solution due to its CMOS compatibility, low power and high-speed operations, and especially high endurance \cite{roypro}. In the STT-MRAM technology, a magnetic tunnel junction (MTJ) is used as a storage element. Information is encoded in a MTJ as a high ($R_{AP}$) or  low ($R_{P}$) resistance state based on the relative magnetization directions of the ferromagnets constituting the device. In read operations of STT-MRAM, tunnel magnetoresistance ratio (TMR) is used to distinguish the $R_{AP}$ and $R_{P}$ states from each other. On the other hand, during the write operations, the switching threshold current ($I_{C}$) of a MTJ is an important parameter determining the current required to easily switch between the $R_{AP}$ and $R_{P}$ states. Low values of $I_{C}$ are obtained while MTJ dimensions are scaled down. However, scaling down the dimensions of MTJ has its own drawbacks. These drawbacks often cause a read disturbance (RD) and incorrect decisions because of low sensing margin (SM) \cite{roypro}. This is because, while low $I_{C}$ is required for low power and low area overhead, it can lead to RD because the minimum reading current ($I_{RD}$) might be close to $I_{C}$. To successfully avoid RD, $I_{RD}$ must be less than $20\%$ of $I_{C}$. The SM issues originate in CMOS technology as a process, supply voltage, and temperature variations, and mainly variations of the oxide thickness ($t_{OX}$) in MTJ technology. In particular, the large TMR will be chosen to increase the SM. Interfacial-anisotropy-type perpendicular MTJ (IPMTJ) have large TMR values such as TMR ratio of $248\%$ \cite{hightmr}, low resistance-area(RA), and high thermal stability. Importantly, IPMTJs consume low switching energy \cite{hightmr}. Note that there is a trade-off between the RD and the SM required $I_{RD}$ \cite{zhaotran1}. A low value of $I_{RD}$ is better for RD rather than the SM. 
The previous works on the sensing scheme of STT-MRAM have highlighted several sensing approaches that are structured in the self-reference cells \cite{selfref}, two transistors and two MTJs cells \cite{twotmtj}, the dynamic data dependent reference cells \cite{covalent},\cite{zhaotran1}, \cite{brief2017} and the only $R_{P}$ reference cells \cite{adequate}. These works are aimed to address the vulnerabilities of a sensing scheme, and especially to increase the low SM. However, using simple current mirror circuitry in the sensing schemes \cite{zhaotran1},\cite{adequate}, \cite{brief2017} can address the vulnerability of a sensing scheme. The dynamic data dependent sensing schemes \cite{covalent}, \cite{zhaotran1}, \cite{brief2017} are a good solution to increase the SM and to decrease bit error rate (BER). However, the power consumption of the sensing scheme will be increased because of utilization of two latch based sense amplifier (SenseAmp) or differential amplifier for $R_{AP}$ and $R_{P}$ states. Although the realization of STT-MRAM cell arrays with two transistor and two MTJs succeeds to increase the low SM, it proves  to be detrimental in terms of low cost and low area \cite{twotmtj}. To emphasize, multiple SenseAmps can be used to decrease readout delay time \cite{nauta}, at the same time to increase the decision circuity (DC) sensitivity, which is composed of cross-coupled inverter pairs in the sensing structure \cite{covalent}. Lowering DC sensitivity will increase the BER. Reducing the secondary noise effects on the DC such as capacitive couplings (CCs) and hysteresis noises improves BER performance. However, investigations have been relatively scarce \cite{kickback} with respect to these secondary noise effects on the DC. In addition most of the works have been focused on improving low SM by increasing the voltage or the current difference between the selected cell and the reference cell. It is important to note that the debate over the offset-aware design of a sensing scheme comes into conflict between a low power design or very precise design because of a trade-off between the power and the offset \cite{nauta}. In this brief, we explore the possible difficulties of a sensing scheme for the STT-MRAM structured with IPMTJs. Meanwhile, we assumed these devices have 1nm oxide thickness ($t_{OX}$) and low resistance-area product RA. The thin ($t_{OX}$) values will increase the BER because of the high resistance variation of the IPMTJ devices. To improve the BER result of the STT-MRAM structured with IPMTJ, we investigated a possible BER performance improvement on the STT-MRAM structured with IPMTJ by increasing the sensitivity of the DC while reducing these noise effects. Also, we review both the voltage mode SenseAmp (VSenseAmp) and the current mode SenseAmp (CSenseAmp) in terms of their decision failure rates (BER). At the same time we compared the VSenseAmp and the CSenseAmp in terms of their high-speed reading operations under high threshold voltage ($V_{TH}$) variations of the transistors in the CMOS deep submicron technologies (i.e., 65m, 45nm). In the point of view, VSenseAmp is robust than CSenseAmp in terms of high speed operation \cite{sinha} when they are designed in these technologies. To clarify, for this comparison, the VSenseAmp is adopted \cite{adequate} and the CSenseAmp is designed similar to \cite{adequate}. In addition, the clamped reference circuitry of the CSenseAmp and the VSenseAmp are not self biased because high resistance variations in IPMTJ devices are not preferred to use as a voltage or current reference source differently than \cite{adequate}. By extension, in our proposed SenseAmps, the reference cells of STT-MRAM are set to the only $R_{P}$ states due to their lower resistance variation than $R_{AP}$ state \cite{adequate}. Moreover, we used three serially stacked IPMTJ expanding their area three times bigger (3S) than their normal size (S) to ensure the RD protection of reference cells with low cost \cite{multiple}. In this approach, without doubt, the increased $I_{C}$ of the reference cells provides the RD protection on these cells \cite{multiple}. Since each design approach has its pros and its cons, we checked the impact of this design approach on the BER results. Because, one of the key ideas of this work is BER performance degradation or improvement with specified techniques. The second is low power STT-MRAM by using IPMTJ devices. The last one is high speed reading operation by reducing the effects of secondary noise sources. Specifically, applying the clamped reference capacitive neutralization technique (CRCNT) to reduce the impacts of CCs on the DC, will be discussed later.
In conclusion, our contributions;
1) We firstly report the sensing schemes for the STT-MRAM structured with IPMTJs. 
2) We attempt speed comparisons taking into account BER results between VSenseAmp and CSenseAmp.
3) We assess the significance of CCs and hysteresis noise effects on BER performance.
In our belief, low power and high-speed sensing scheme design efforts will be pursued with the renewed size dimensions of the MTJ devices decreasing below 20nm. In fact, it is impossible to cover all specifications in a singular design approach.
In this brief, the rest of the paper is prepared as follows: Section II studies the design aspects of the proposed VSenseAmp, Section III shows the results and their effects on the key design metrics, Section IV concludes this brief.
\begin{figure}[ht!]
\centering
\includegraphics[width=3.0in,keepaspectratio]{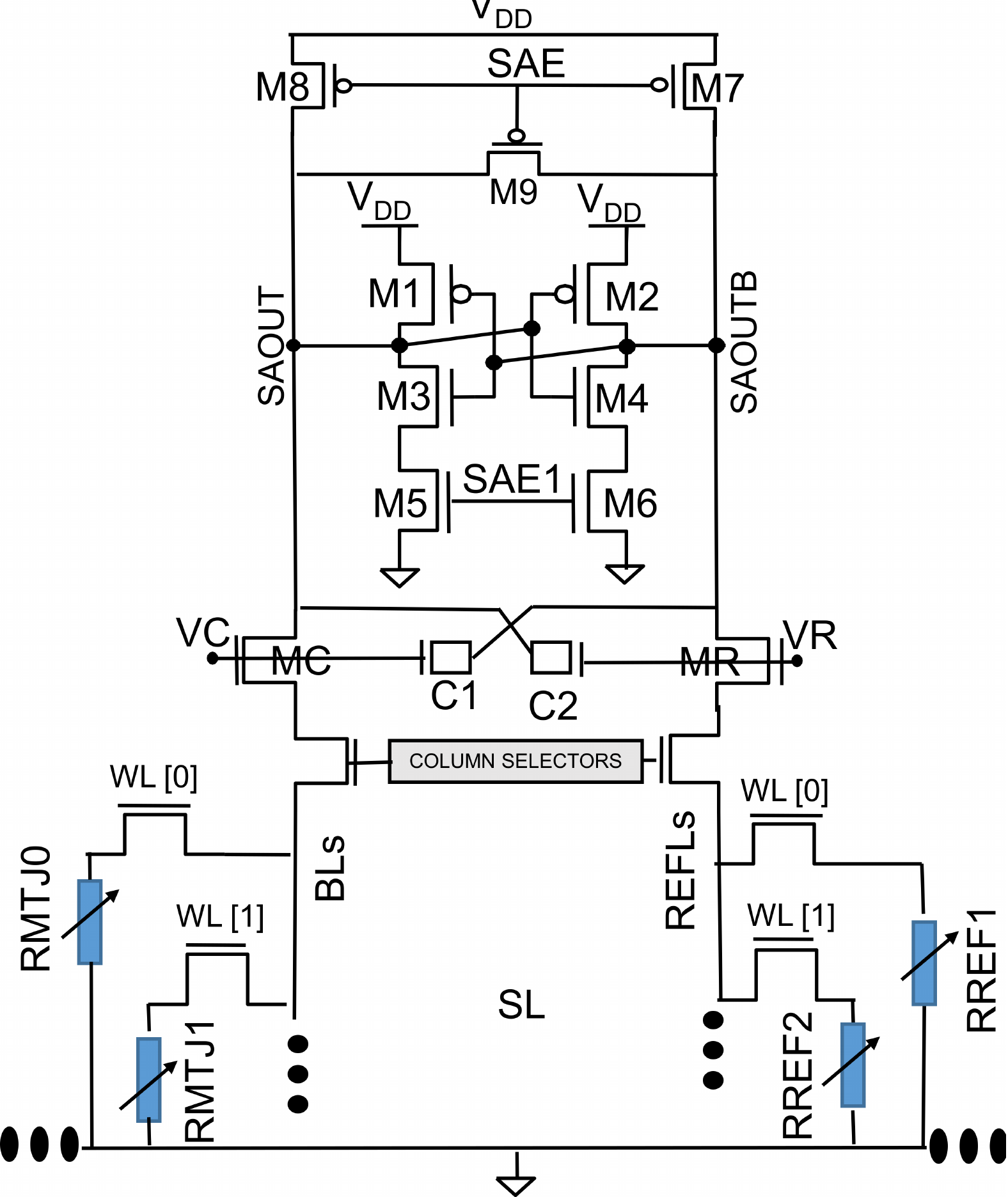}
\caption{The schematics and array structure of the proposed NVSenseAmp.}
\label{fig:fulcirc} 
\end{figure}

\section{THE EVOLUTION OF THE NVSenseAmp}
A SenseAmp in STT-MRAM converts resistive information to digital information during reading phases. A latch type SenseAmp is widely used due to its positive feedback mechanism. Reading phases in the proposed SenseAmps are carried out in three phase: pre-charge, evaluation, and decision. Our timing strategy for the proposed SenseAmps is adopted \cite{adequate} and our clock signals are SAE, SAE1, and WL. In the pre-charge phase, the memory cells or the reference cells are activated through word lines (WLs), and then both BLs and REFLs are pre-charged to $V_{DD}$ equalizing via $M_{9}$ to provide same delays among the BLs and the REFLs. In the evaluation phase, the proposed SenseAmps are activated with SAE and disconnected from cell arrays via SAE1. And then seen a differential voltage at the output nodes of the proposed SenseAmps (SAOUT, SAOUTB). Finally, this differential output voltage is amplified by the DC via its high gain positive feedback mechanism and then converted to digital signal. 
A VSenseAmp takes more time than a CSenseAmp because of longer discharging time of BL capacitance ($C_{BL}$) or REFL capacitance ($C_{REFL}$). In particular, the readout speed of the VSenseAmp is faster than the CSenseAmp when the variation of $V_{TH}$ is higher than 12mV \cite{sinha}. The variations of $V_{TH}$ are 30mV or more in deep submicron technologies. In our simulations, we used TSMC 65nm CMOS model parameters and the MTJ model from \cite{knack}. Particularly, some essential parameters are listed as follow: power supply ($V_{DD}$) is 1.0V, the dimensions of IPMTJ are 40(width)x 40(length)x 1(oxide thickness)$ nm^{3}$. In addition, the process variations of MTJs and CMOS are used $2\%$ ($3\sigma$), and specified using TSMC 65nm CMOS model. Our proposed VSenseAmp with CRCNT, that is, the NVSenseAmp with only $R_{P}$ reference cells, as shown in Fig.1. 
\begin{figure}[ht!]
\centering
\includegraphics[width=3.0in,keepaspectratio]{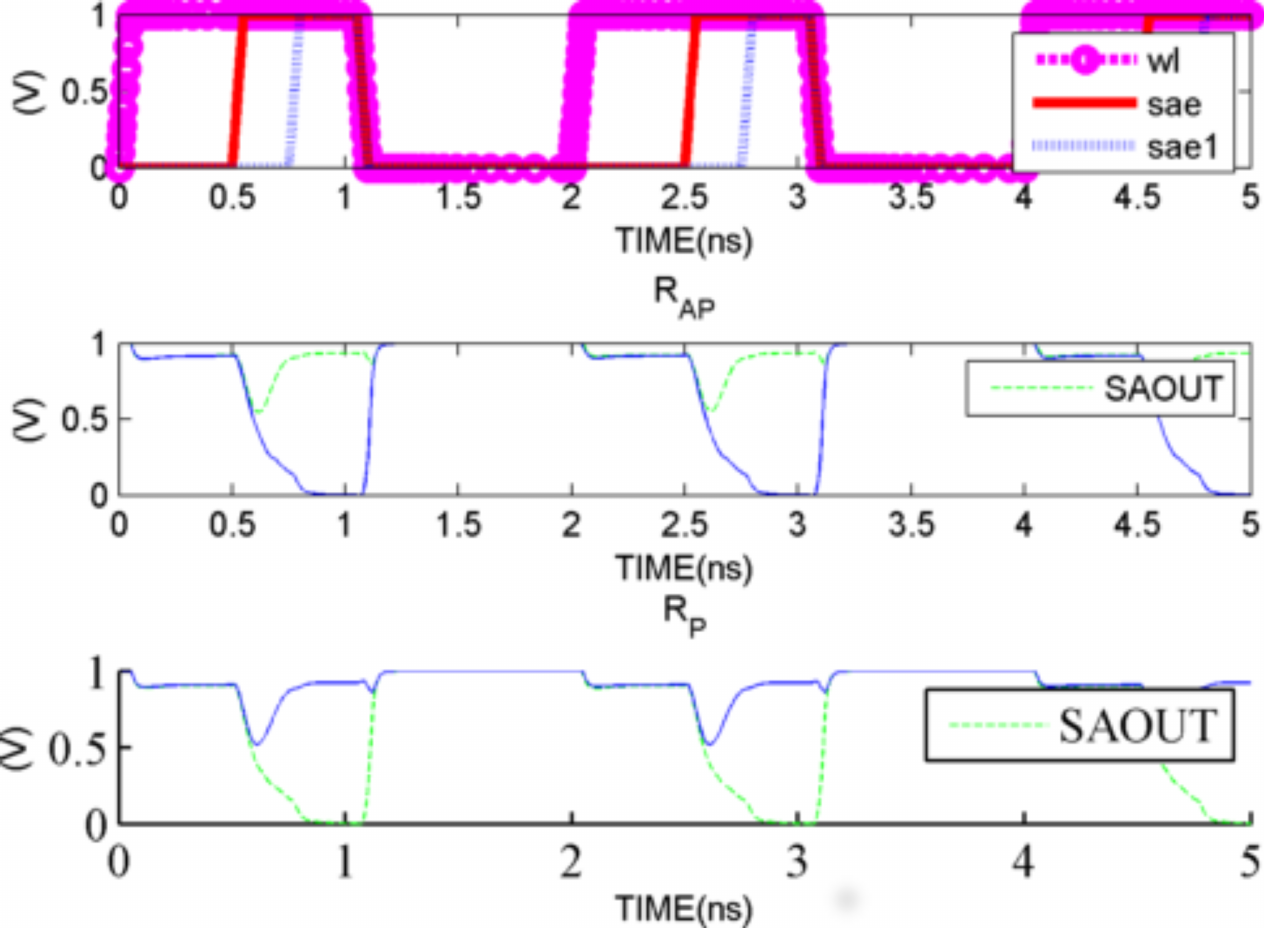}
\caption{The timing diagrams of the proposed SenseAmps.}
\label{fig:2} 
\end{figure}
The activation orders for the stages in the proposed SenseAmps are shown in Fig.2. Firstly, the WL is activated at 0ns with 1ns of pulse width. Secondly, the SAE is activated at 0.5ns with 0.5ns of pulse width. Thirdly, the SAE1 is activated at 0.75ns with 0.25ns of pulse width. The output signals of the NVSenseAmp for the $R_{AP}$ and $R_{P}$ states are shown in Fig.2. 
\subsection{The Hysteresis Effects}
The sensed data which is previously stored in parasitic capacitances, mainly $C_{GD}$) of the SAOUT and SAOUTB nodes of the proposed SenseAmps, and this previous stored data causes hysteresis effects. Furthermore the hysteresis effects causes the sensing errors when the recovery time of the DC is inefficient. As a solution, we propose a technique which the SAOUT and SAOUTB nodes are directly connected to the drains of MC and MR (the clamped reference transistors) to reduce the effects of $C_{GD}$ on these nodes, respectively. We tested the recovery time of the NVSenseAmp. Regarding the test configuration, we assumed that the load capacitances of NVSenseAmp are initially set at 0V or $V_{DD}$ to stimulate these effects on the SAOUT and SAOUTB nodes. As a result, we obtained the same BER or failure rates, for the NVSenseAmp, by taking into account the hysteresis effects.
\subsection{The Capacitive Couplings Effects on the Clamped Reference}
The main function of clamped reference circuitry (MC and MR) is that it provides both overcurrent protection and an inequality between BLs and REFLs to easily distinguish a $R_{P}$ state when the data and the reference cells are the $R_{P}$ state. Meanwhile, the MC and MR are biased at $VC=0.8$ and $VR=0.7$ values that are chosen for the optimal BER via the parametric analyzes. However, these clamped reference transistors will be driven in the deep triode region to optimize the BER, so that the values of VC and VR can be chosen as high as possible. The output nodes (SAOUT and SAOUTB) of SenseAmps have large output swings. These variations cause a voltage disturbance in the gate of the clamped reference transistors (MR and MC) and current accuracy errors in the read current through CCs. The main CCs on the DC of the CSenseAmp and the NVSenseAmp are shown in Fig.3a and Fig.3b, respectively. CRCNT is realized through C1 and C2 MOS capacitances to minimize the effects of CCs, as shown in Fig.1. The CRCNT works  as a capacitive voltage divider between a BL and a REFL to balance the read and the reference cell currents. In addition, the values of $C_{GD}$ vary the width and length of a MOS transistor. Therefore the size of the MOS capacitances (C1 and C2) should be chosen the same as MC and MR. It is important to note that currents of BLs and REFLs can be balanced together with these capacitances, as formulated in Eq.1-2. Indeed, The reduced effects of CCs can boost the high-speed operation of the NVSenseAmp.
\begin{equation}
\frac{V_{SAOUT}-VC}{V_{SAOUT}-V_{SAOUTB}}= \frac{C_{GDC}}{C_{1}+C_{GDR}} 
\end{equation}
\begin{equation}
\frac{V_{SAOUTB}-VR}{V_{SAOUTB}-V_{SAOUT}}= \frac{C_{GDR}}{C_{2}+C_{GDC}}
\end{equation}

\begin{figure}[ht!]
\centering
\includegraphics[width=2.5in,keepaspectratio]{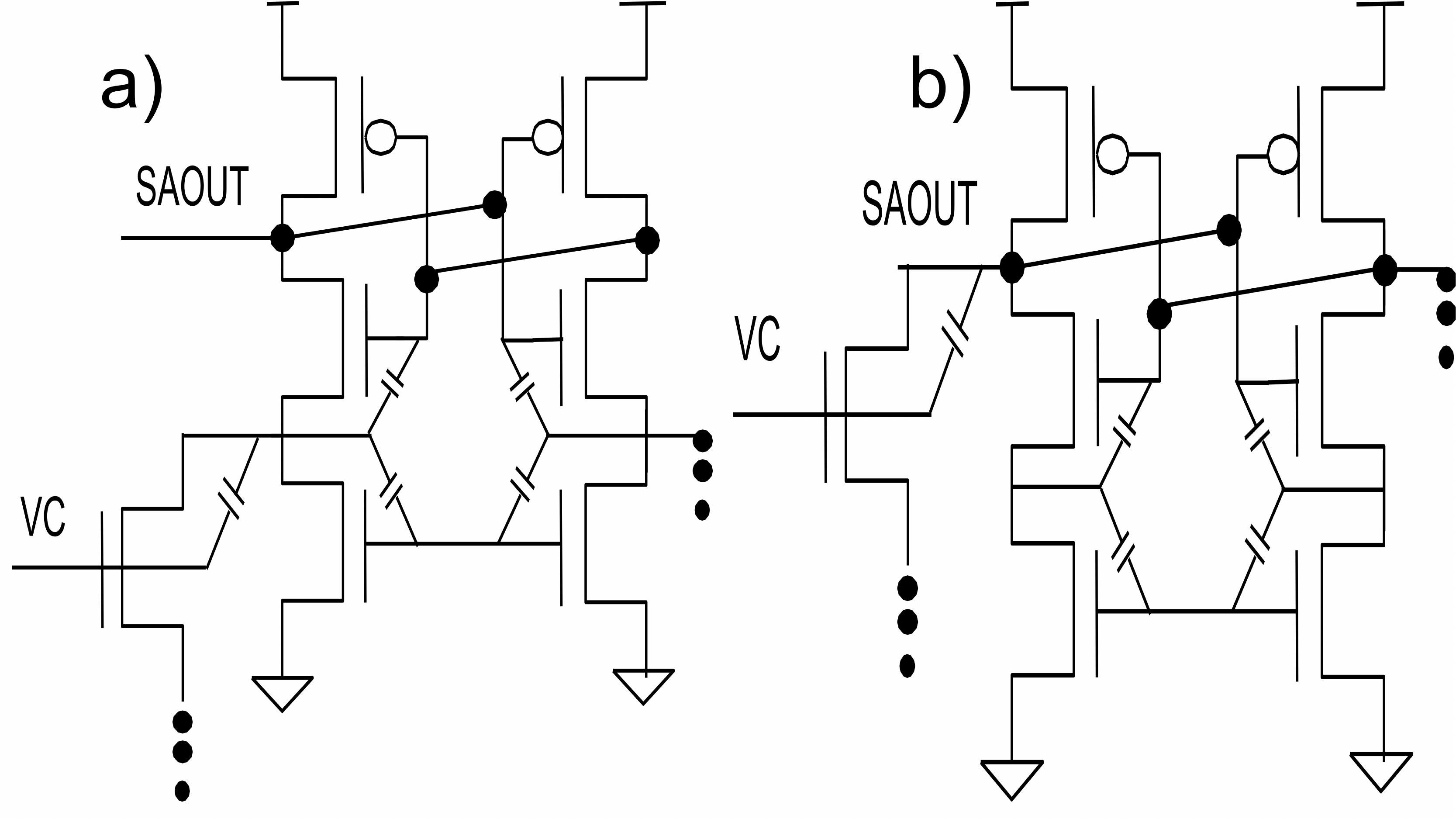}
\caption{The left side of DC on a) the CSenseAmp b) the NVSenseAmp.}
\label{fig:kickexp} 
\end{figure}

\subsection{Power Comparisons}
Our proposed SenseAmps are structured dynamically. The power dissipation is  mainly determined by the duty cycle and voltage swings on the BLs and REFLs, as formulated in Eq.3.  
\begin{equation}
P_{dyn}=\alpha (activity)x C_{total} x V_{swing} x V_{DD}x f
\end{equation}
Firstly, the DC is a power hungry unit needs to effectively disconnected during read cycles to advance the power performance of the proposed SenseAmps. The disconnection of the DC is controlled by SAE1. We analyzed possible timing strategies of SAE1 as given following approaches: First, SAE1 is taken the same as SAE, second SAE1 is taken after the SAE with one or two inverter delays. However, we did not find the improved solution different than \cite{adequate}. Thirdly, we applied CRCNT to reduce the effects of parasitic capacitances. 
The reading current of the CSenseAmp is less than both the VSenseAmp and the NVSenseAmp. The NVSenseAmp or the VSenseAmp with the multiple reference cells in the $R_{P}$ state have almost the same power dissipation as they are configured with single reference cell in the $R_{P}$ state, indicated in Table I. As a result, the power dissipation is the lowest in the CSenseAmp, and the highest in the NVSenseAmp. However, the output voltage of the NVSenseAmp has not rail to rail voltage swings, so will be provided by taking the outputs of the NVSenseAmp after an inverter stage but the given power results of the NVSenseAmp will be increased. 
\begin{table}
\renewcommand{\arraystretch}{1.3}
\caption{The Power($\mu W$)Consumption Comparisons}
\label{table_3}
\centering
\begin{tabular}{c c c}
\hline \hline
\bfseries { Designs} & \bfseries {$R_{P}$}& \bfseries { $R_{AP}$}\\
\hline
{The CSenseAmp with single reference cell} &32  &29  \\
{The VSenseAmp with single reference cell} &45 &38 \\
{The VSenseAmp with multiple reference cells} & 45 &38  \\
{The NVSenseAmp with multiple reference cells} &48 &41  \\
{The NVSenseAmp with single reference cell } &48  &41  \\
\hline \hline
\end{tabular}
\end{table}
We compared the power dissipation (at 66.7MHz) of the some previous works and the NVSenseAmp. Our proposed design has less power dissipation than these previous works, as shown in Fig.4. In addition, our power dissipation results are obtained through Monte Carlo simulations (with 1K samples), and separately for the $R_{AP}$ and $R_{P}$ states.
\begin{figure}[ht!]
\centering
\includegraphics[width=2.6in,keepaspectratio]{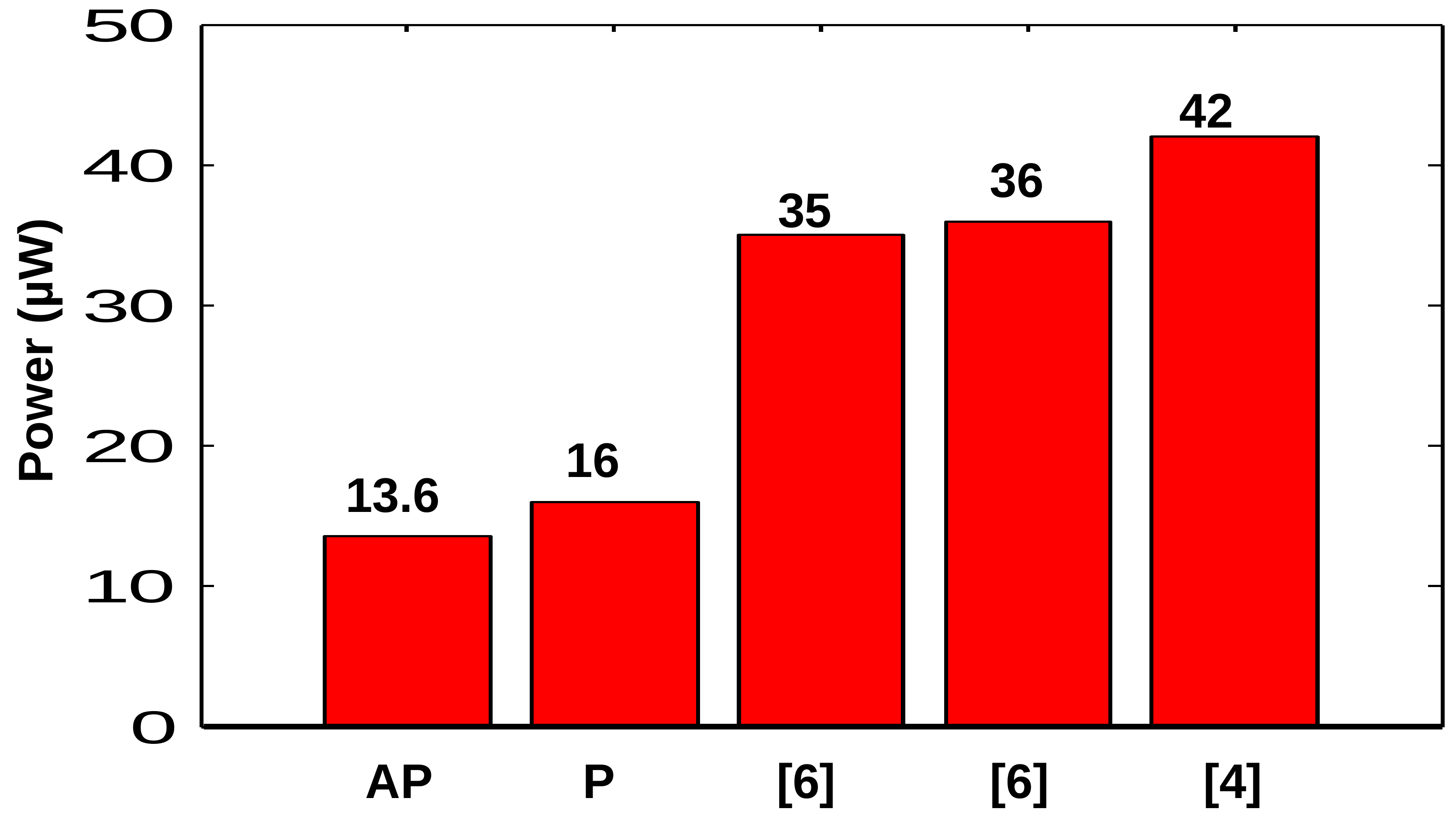}
\caption{The average power comparison of the previous works and our proposed NVSenseAmp.}
\label{fig:power} 
\end{figure}

\subsection{Readout Time Comparisons}
The clamped bitline scheme can be used to reduce the readout delay of the proposed sensing schemes. The readout time of CSenseAmp is insensitive to $C_{BL}$. For this reason, in our simulations, we compared the readout time between the proposed SenseAmps for 50fF which is the predicted maximum $C_{BL}$ in sub memory cell arrays. Basically, the readout time depends on the transconductance of nmos transistors in DC and the load and parasitic capacitances of DC. Secondarily, the readout time of VSenseAmp is sensitive to the discharging time of $C_{BL}$ and the voltage swing of BLs, on the contrary CSenseAmp. As a comparison, the NVSenseAmp and the VSenseAmp are faster than the CSenseAmp because of above 12mv threshold voltage ($V_{TH}$) values \cite{sinha} in deep submicron technologies (in this design, 65nm). The NVSenseAmp is faster than the VSenseAmp with a single reference but slower than the VSenseAmp with multiple references  because of the increased read current values of BLs. Importantly, the NVSenseAmp is less sensitive to the increased current of BLs due to CRCNT. However, these readout delay comparisons are not the same for the $R_{AP}$ and $R_{P}$ because of the asymmetric resistance distribution of $R_{AP}$ and $R_{P}$. In addition, the parasitic output capacitance value considering the total of $C_{GD}$ of the NVSenseAmp is greater than the CSenseAmp. As a reminder, the high speed operation will consume high power in terms of speed-power trade-off. As a result Table II shows the comparison of the readout time of the proposed designs. 
\begin{table}
\renewcommand{\arraystretch}{1.3}
\caption{Readout Time(ps) Comparisons}
\label{table_2}
\centering
\begin{tabular}{c c c}
\hline \hline
\bfseries { Designs} & \bfseries {$R_{P}$}& \bfseries { $R_{AP}$}\\
\hline
{The CSenseAmp with single reference cell} &656  &707  \\
{The VSenseAmp with single reference cell} &669 &649 \\
{The VSenseAmp with multiple reference cells} & 167 &151  \\
{The NVSenseAmp with multiple reference cells} &271 & 244 \\
{The NVSenseAmp with single reference cell} &270  &244  \\
\hline \hline
\end{tabular}
\end{table}
Fig. 5 shows the readout time comparisons of the previous works and the NVSenseAmp, our readout time value is a mean value obtained from Monte Carlo analysis (1K samples). As a conclusion, the NVSenseAmp has faster readout time among these compared works. 

\begin{figure}[ht!]
\centering
\includegraphics[width=2.6in,keepaspectratio]{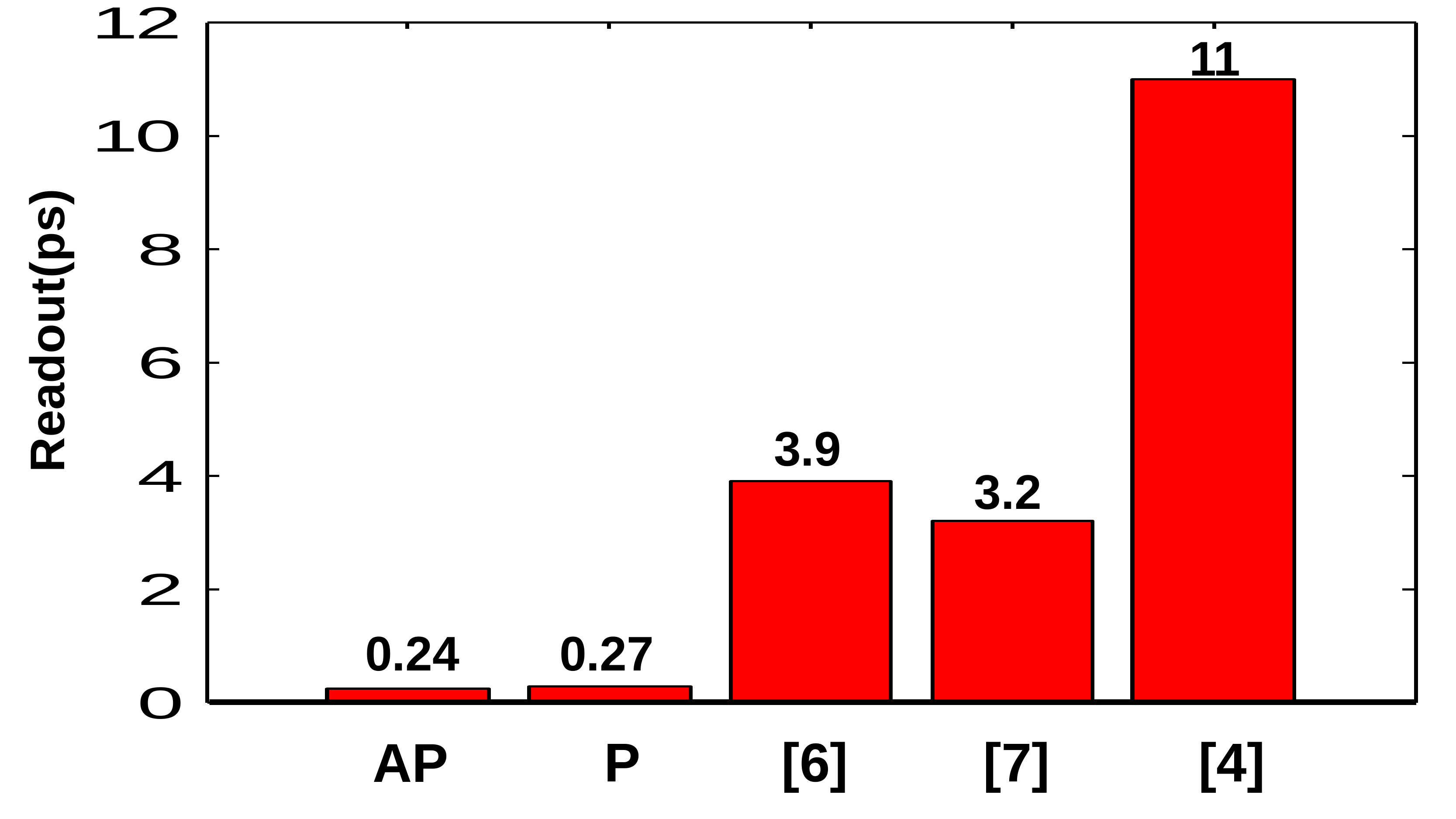}
\caption{The readout delay times comparison of previous works with the proposed design.}
\label{fig:speed} 
\end{figure}

\subsection{The Resistance Variations}
The resistance variations of the IPMTJs in our designs are $13\%$ for both $R_{P}$ (the current value is 742$\Omega$) and $R_{AP}$ (the current value is 1.97K$\Omega$) associated with $ t_{OX}$ (1nm) variations of $2\%$ for 3$\sigma$. These resistance variations are more than the variations of which are commonly taken as $5\%$ and 3$\sigma$ in previous works. Figure 6 shows the mean variations of $R_{AP}$ and $R_{P}$. Comparatively, their variations are close to each other despite their asymmetric structures.
\begin{figure}[ht!]
\centering
\includegraphics[width=2.8in,keepaspectratio]{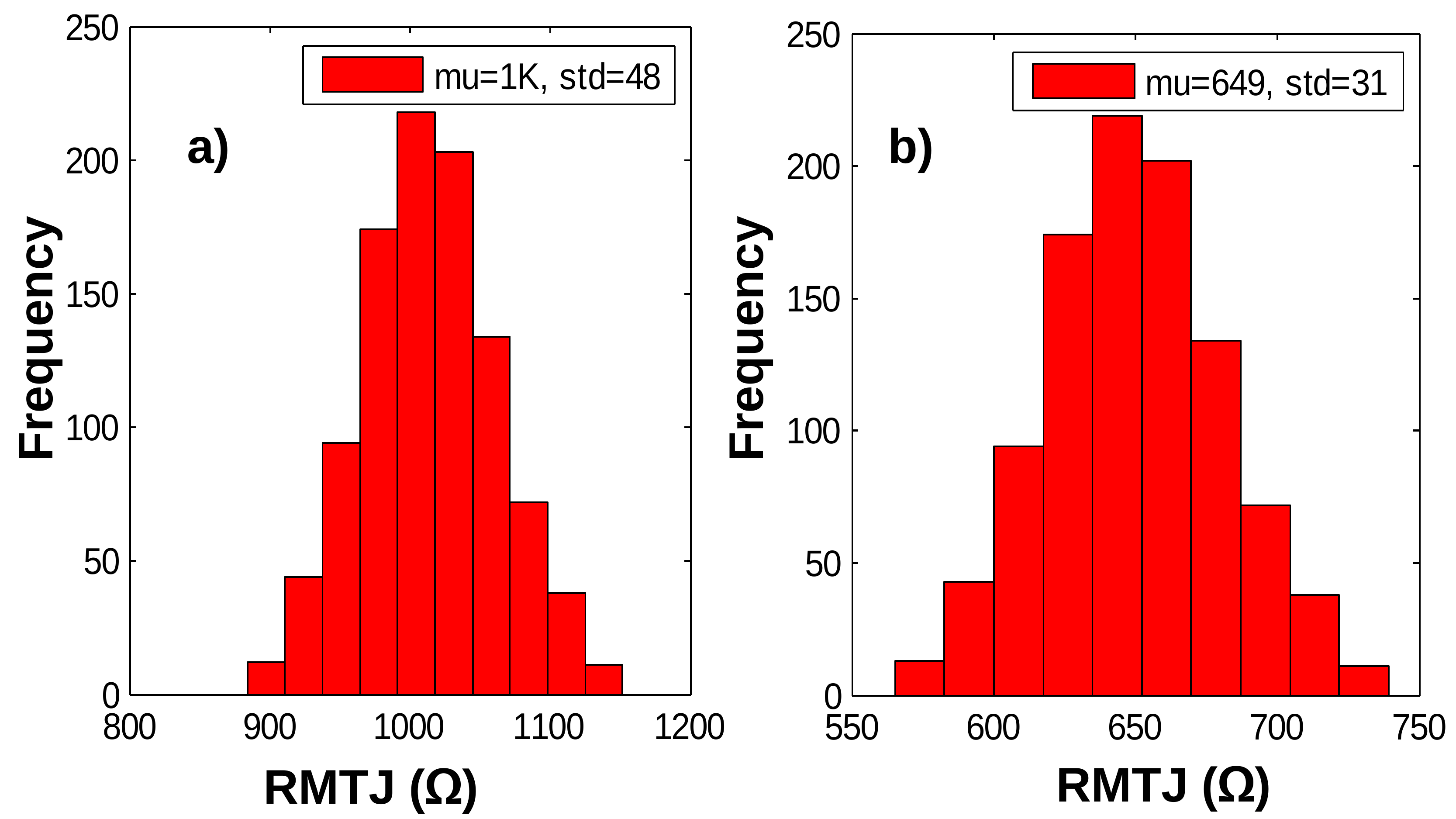}
\caption{The resistance variations of the proposed IPMTJ cell for a) The AP state b) The P state.}
\label{fig:resvar} 
\end{figure}

\subsection{The BER Comparisons}
The effects of CMOS and MTJ process variations for the BER performance of the proposed SenseAmps were evaluated through Monte Carlo simulations with 1K samples. IPMTJ devices have high TMR values but the resistances of the IPMTJs are comparable with the resistance of the access transistors. However, the higher TMR values of IPMTJs will be better for SM and BER results \cite{zhaotran1}.  To emphasize that the IPMTJ devices have the high TMR values with the values of $t_{OX}>0.8nm$ but the TMR values change exponentially below the values of $t_{OX}<0.8nm$ \cite{wang2013}. As a matter of fact our opinion is that difficult to find a optimum solution in terms of BER performance between $R_{P}$ and $R_{AP}$ states, as shown Table III. By comparison the CSenseAmp has lower BER performance for the $R_{P}$ state than the VSenseAmp or the NVSenseAmp, shown in Fig.7-8. The NVSenseAmp has the partially better BER performance among the other proposed SenseAmps.   
\begin{table}
\renewcommand{\arraystretch}{1.3}
\caption{BER Comparisons}
\label{table_1}
\centering
\begin{tabular}{c c c}
\hline \hline
\bfseries {Designs} & \bfseries {$R_{P}$}& \bfseries { $R_{AP}$}\\
\hline
{The CSenseAmp with single reference cell} &427 &0  \\
{The VSenseAmp with single reference cell} &17 &4 \\
{The VSenseAmp with multiple reference cells} & 17 & 6  \\
{The NVSenseAmp with multiple reference cells } &8 &6  \\
{The NVSenseAmp with single reference cell} &2 &8 \\
\hline \hline
\end{tabular}
\end{table}

\begin{figure}[ht!]
\centering
\includegraphics[width=2.5in,keepaspectratio]{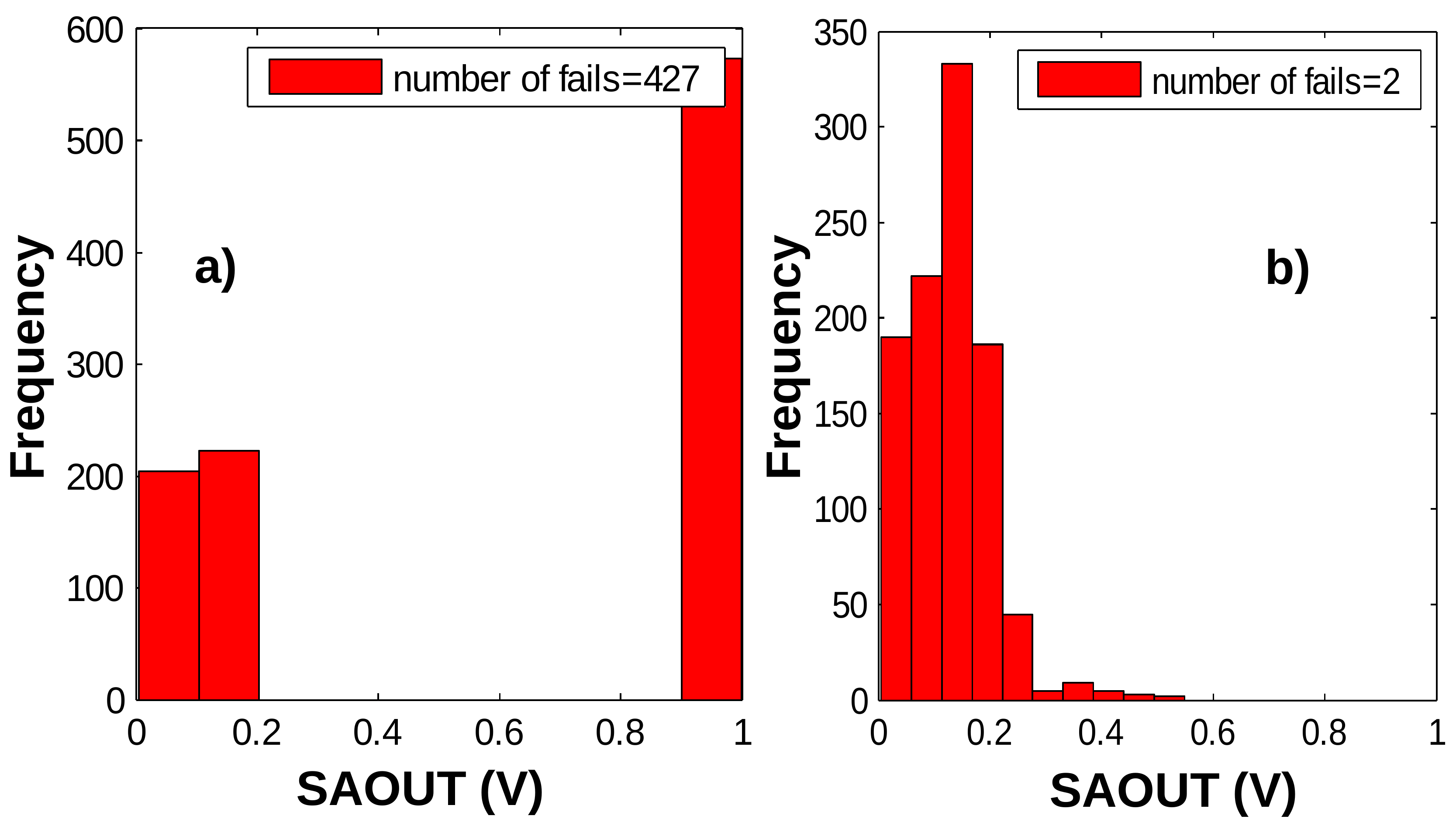}
\caption{Monte Carlo Simulations for P states a) The CSenseAmp b) The NVSenseAmp.}
\label{fig:p} 
\end{figure}

\begin{figure}[ht!]
\centering
\includegraphics[width=2.5in,keepaspectratio]{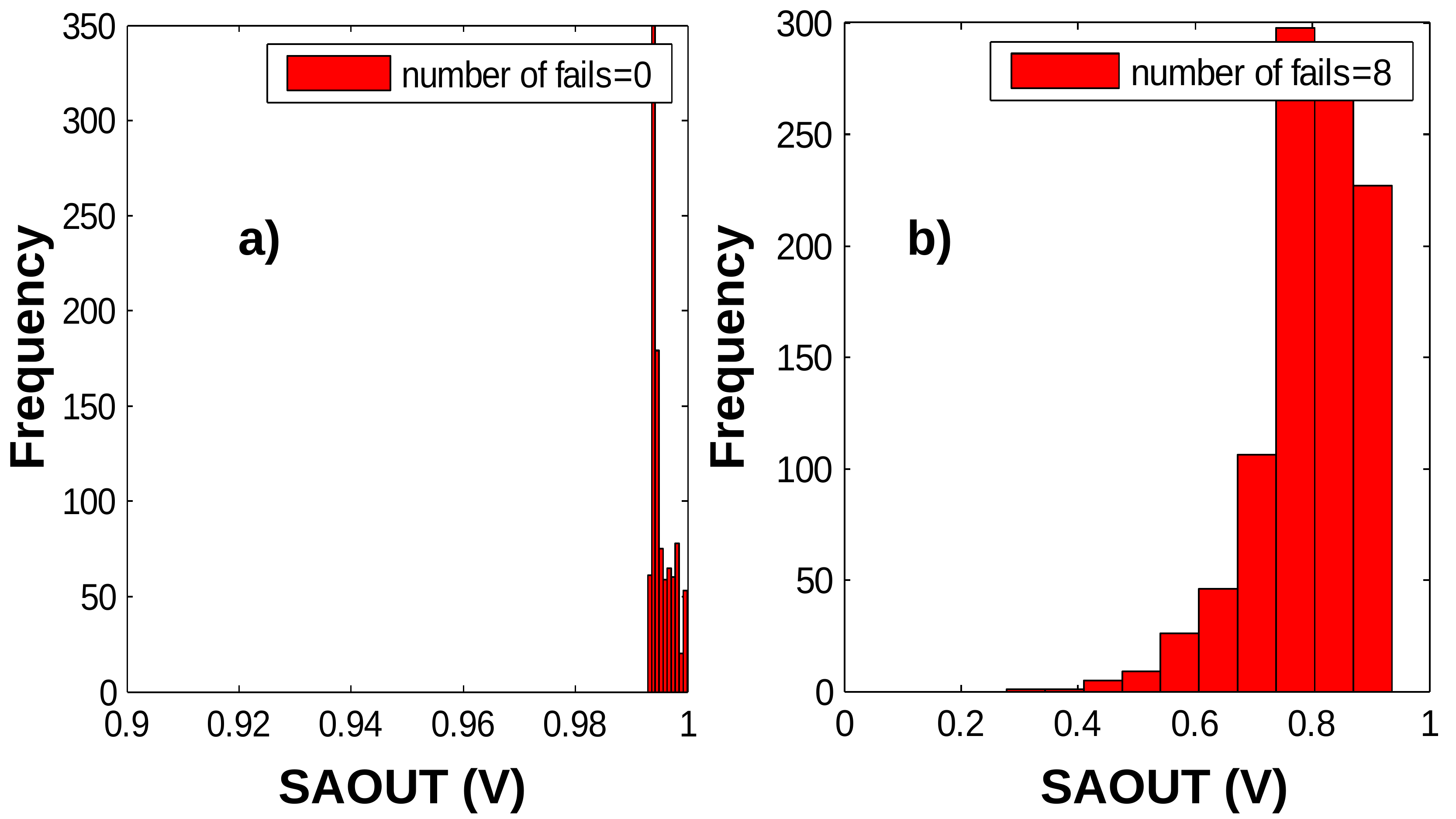}
\caption{Monte Carlo Simulations for P states a) The CSenseAmp b) The NVSenseAmp.}
\label{fig:ap} 
\end{figure}

\section{Conclusion And Discussions}
In this brief, we investigated the evolution of sensing schemes for STT-MRAM structured with IPMTJs. Moreover, we compared the proposed sensing schemes in terms of power, speed, and BER results. As a comparison the NVSenseAmp has a great advantage in sensing speed and a partial better BER performance among the other proposed designs that are the CSenseAMP and the VSenseAmp. However, the design of the NVSenseAmp or the VSenseAmp with multiple reference cells is a valuable solution for RD protection but degrades the BER performance of the NVSenseAmp or the VSenseAmp. Consequently, the proposed NVSenseAmp with single reference cell is a good solution for high-speed and low-power reading operations among the compared literature works. To enumerate the read operations of the NVSenseAmp have successfully been performed a 2.5X reduction in average low power and a 13X increase in average high speed compared with the previous works. STT-MRAM structured with IPMTJs as a remarkable candidate as a universal memory have its low switching energy. However, their accuracy rates are low and must be increased by developing new techniques. Our future work will be a design of offset reduction technique for the NVSenseAmp to improve the accuracy rates of NVSenseAmp.
\section*{Acknowledgment}
This work is part of a project that has received funding from the European Union's H2020 research and innovation programme under the Marie Skłodowska-Curie grant agreement No 691178, and by the TUBITAK-BIDEB 2214/A.
\ifCLASSOPTIONcaptionsoff
  \newpage
\fi
\bibliographystyle{IEEEtran}
\bibliography{senseref}
\end{document}